# TREATING CROWDSOURCING AS EXAMINATION: HOW TO SCORE TASKS AND ONLINE WORKERS?

Guangyang Han[1] and Sufang Li[2] and Runmin Wang[3] and Chunming Wu[4]

[1, 2, 3, 4]College of Computer and Information Sciences, Southwest University, Chongqing, China

[1, 2, 3][hanguangyang, sufangli, rmwang]@email.swu.edu.cn
[4]springsun@swu.edu.cn

## ABSTRACT

*Crowdsourcing is an online outsourcing mode which can solve the machine learning algorithm's urge need for massive labeled data. How to model the interaction between workers and tasks is a hot spot. we try to model workers as four types based on their ability and divide tasks into hard, medium and easy according difficulty. We believe that even experts struggle with difficult tasks while sloppy workers can get easy tasks right. So, good examination tasks should have moderate degree of difficulty and discriminability to score workers more objectively. Thus, we first score workers' ability mainly on the medium difficult tasks. A probability graph model is adopted to simulate the task execution process, and an iterative method is adopted to calculate and update the ground truth, the ability of workers and the difficulty of the task. We verify the effectiveness of our algorithm both in simulated and real crowdsourcing scenes.*

## 1. INTRODUCTION

With the rise of the Internet in the early 21st century, many traditional industries are undergoing changes, such is outsourcing. The first clear definition of crowdsourcing came in 2006, when Howe [1] defined it as: "Simple defined, crowdsourcing represents the act of a company or institution taking a function once performed by employees and Outsourcing it to an undefined (and generally large) network of people in the form of an open call."

Early crowdsourcing work mainly come from the field of database. Researchers package crowdsourcing platform and workers into a virtual database, and query some open or abstract questions through SQL-like query language. These queries are eventually transformed into crowdsourcing tasks, which are completed by a large number of Internet workers and fed back to the inquisitors. In recent years, with the rise of machine learning, especially deep neural network, researchers have an increasing demand for massive labeled data, and crowdsourcing has also attracted a lot of attention from data mining and machine learning field. The famous *ImageNet* dataset [2] was crowdsourced by Google through *Amazon Mechanical Turk* (*AMT*), a general crowdsourcing platform. There are other online platforms like *FigureEight* (formerly named as *CrowdFlower*), *InnoCentive*, *Upwork*, *etc*.

Crowdsourcing has various task patterns. For example, according to the task granularity, crowdsourcing can be divided into micro and macro task, such as annotating an image and polishing a paper. Also, crowdsourcing could be classified into monetary, entertainment or voluntary services in terms of the incentive mechanism. *Foldit* [3] is an online game in which people match to optimize the 3d shape design of proteins, making people in the process of game also can contribute for biological sciences, and *Wikipedia*, for example, is enriched and maintained by volunteers. More, based on the purpose of crowdsourcing, individuals, collaborate and competition are three participation mode for workers in crowdsourcing. People solve independent sub-problems such as image annotation in parallel, collaborate on complex tasks such as writing a program system serially, and compete to win a creative design competition.

Crowdsourcing tasks that provide labeled data for machine learning include many forms, such as data collection, categorization, transcription, or relevance searching and sentiment analysis [4] , *etc*. Our main focus is the data annotation task, which categorizes and annotates unlabeled data. That is, tasks contain several options from which workers choose the most appropriate one. The general process for crowdsourcing can be roughly divided into three steps: requester designs and submits tasks to the crowdsourcing platform, the platform assigns tasks to appropriate workers, then collects and aggregates workers' answers and feeds back to the requester [5]. All three stages contribute to the quality of the final result. In a word, clear and intuitive task design, reasonable task assignment strategy and appropriate truth inference algorithm are the three hot spots of crowdsourcing research [6], furthermore, the last two stages all rely on pertinent worker and/or task model that can reveal the real situation.

We get the inspiration for modeling workers and tasks from school exams. There are two basic indicators in the evaluation of test questions, difficulty and discriminability. Difficulty corresponds to the overall scoring rate (correct rate) of the question. A good question should be able to distinguish between students of different levels of knowledge, that is, the high scoring rate of the good student and the low scoring rate of the student who does not work hard. The difference between the two evaluating indicator is the degree of discriminability. In general, the discriminability is highest when the problem is of moderate difficulty. In this way, we should pay more attention to medium difficult tasks when inferring workers' abilities from task results, since tasks that are too difficult or too easy contain little information about workers' abilities.

In general, we use two positive decimals to quantify task difficulty and worker ability, a probability graph model to simulate the annotating process, and comprehensively consider the impact of difficulty and ability on worker's performance in the deduction process. In particular, when inferring worker ability from the annotating results, we focus more on the performance on moderate difficulty tasks, reducing the impact of simple and difficult tasks, thus making our model of worker competence more accurate and robust. Our innovations are included as the following:
1.  We draw inspiration from actual exam, using difficulty to measure tasks, and ability to measure workers. A probability graph model simulates the annotating process, which is simple and intuitive.
2.  We *first* point out the effect of task discriminability on inferring workers' ability, and design a set of computational inference process to infer more accurate and robust worker model.
3.  We designed simulation experiment and real experiment to verify the effectiveness and practicability of our algorithm.

## 2. RELATED WORK

Data mining in crowdsourcing focuses more on truth inference algorithms, and good raw data is indispensable. However, with limited budget, we need to select some from a large number of not so reliable online workers to complete our tasks, so task assignment has also attracted a lot of research work. In order to better assign tasks or infer ground truth, we need to have a full understanding of workers and tasks, thus the modeling of workers and/or tasks is also inevitable. The modeling of workers mostly focuses on their ability level, the difference lies in different granularity, such as overall ability, or abilities in different domains, and even use matrix to express the cognitive preference of workers in various domains/options. For task modeling, besides focusing on the nature of the task itself, such as difficulty or domain, some researchers also measure the degree of completion of a task, such as uncertainty, information entropy, and so on.

## 2.1. Task Assignment

The core of task assignment is how to allocate tasks to appropriate workers so as to maximize the overall return [7]. Workers should have enough ability and even interest to complete the tasks assigned to them. According to whether it is assumed to know the global information of workers and tasks, such as skill level, difficulty, domain, price, *etc.*, task assignment algorithms can be divided into online and offline types. In the real crowdsourcing process, the information of workers and tasks is usually unknowable at the initial stage, so we mainly focus on online algorithms.

In the online algorithms, workers actively access the platform, and the platform has no prior information of the workers initially. The requester provides tasks to the platform, which allocates tasks when the workers arrive. Due to the lack of knowledge of the unknown workers' abilities, the assignment may be random at the beginning, but after workers returning their answers, platform can infer and learn stage by stage. With the increase of workers' answers, the platform can gradually grasp the characteristics of the workers and/or tasks, so that the subsequent task assignments can be targeted.

As far as we concern, there are the following typical work that focus on task assignment algorithms:
1. *DTA* [7]: It's a two-stage exploration-development assignment algorithm. In the exploratory stage, they used sampling methods to evaluate the ability of workers and the difficulty of the tasks. After obtaining enough information, the problem was transformed into the offline style, they further used the *Primal-Dual Formulation* to solve it.
2. *AskIt* [8]: A real-time application system for interactive crowd-data searching, which can effectively route questions to the due staff, and reduce the uncertainty of answers. The algorithm considers four constraints and two uncertainty measures of the workers and questions, and is used to solve the optimization problem. Entropy-like methods are used to quantify the uncertainty, while collaborative filtering is employed to predict the unseen answers.
3. *QASCA* [9]: It is an online task assignment framework that uses a confusion matrix to simulate workers' preferences when giving answers, and uses Accuracy and F-score to measure the quality of task completion. It gradually infers the truth value of the task and updates the worker confusion matrix, thereby dynamically assigning the task to the worker with the highest probability of giving the correct answer in the next assignment.

## 2.2. Truth Inference

Worker and task model are also crucial in the truth inference (answer aggregation) stages. Because building these models requires the ground truth of the tasks, which are unknown, many algorithms iteratively perform truth inference and model update calculation.

The truth inference algorithm can be divided according to whether it needs to be calculated iteratively. Non-iterative aggregation uses heuristics to calculate a single result for each task. Commonly used techniques include weighted majority voting (*WMV*) and filter-based techniques [10]. (Weighted) Majority voting gives all answers the same (different) weight, then counts, and the majority wins [11]. Filter-based methods rely on other techniques (such as qualification test) to filter out unreliable answers first, and then perform majority voting. Typical works are *HP* [12], *ELICE* [23], *BV* [5], *etc.*

The iterative aggregation algorithm performs a series of iterations and finally produces high-quality results. It usually consists of two stages: truth inference and model update. The most typical representative is the Expectation-Maximization (*EM*) algorithm [13]. In the E-stage, the truth value of the tasks is inferred based on the existing worker model, and in the M-stage, the model parameters are adjusted to maximize the occurrence probability of the current inferred result. The performance of the iterative algorithm is greatly affected by the initial parameters of the model, and is usually better than the non-iterative methods. However, the iterative process will cost a lot of time and computing resources. You can refer to the following works: *SLME* [14], *GLAD* [15], *ITER* [16], *etc*.

## 3. PROPOSED METHOD

We will elaborate on our proposed model and method in this section. First of all, let's introduce some common terms in crowdsourcing and the notations in this paper.

### 3.1. Terms and Notations

*People* with crowdsourcing needs will hire a group of *workers* to work for their *task* through a *website*. We call them *requester*, crowdsourcing *workers* $\mathcal{W} = \{w_1, w_2, \ldots, w_m\}$, crowdsourcing *tasks* $\mathcal{T} = \{t^1, t^2, \ldots, t^n\}$ and crowdsourcing *platform* respectively. The term task refers to different granularity according to the context, such as the entire task, a single sub-task or a batch of sub-task, so does the term worker. Each task $t^i$ has a potential ground truth label $l^i$, and may receive some annotations $A^i = \{a_1^i, a_3^i, a_6^i, \ldots\}$ from several workers $\{w_1, w_3, w_6, \ldots\}$. At the same time, worker $w_j$ may give out annotation set $A_j = \{a_j^2, a_j^4, a_j^9, \ldots\}$ to tasks $\{t^2, t^4, t^9, \ldots\}$. Please pay attention to the correspondence between serial numbers and subscripts in these two examples. In this paper, we prefer to use subscripts (letter $j$) to distinguish all $m$ workers and superscripts (letter $i$) to mark total $n$ tasks, a task may receive annotations from multiple (not all) workers, and a worker may also annotate multiple (not all) tasks. Due to the subjective or objective factors of tasks and workers, those annotations are not so reliable, we often need to annotate a task repeatedly to obtain more reliable results [17, 18]. We assume that each task gets $r$ annotations on average, which is also the reason for the existence of truth inference algorithms. Truth inference algorithm is used to infer the ground truth set $\mathcal{L} = \{l^1, l^2, \ldots, l^n\}$ from annotation set $\mathcal{A} = \sum_{i=1}^{n} A^i$ (or note as $\mathcal{A} = \sum_{j=1}^{m} A_j$).

As mentioned earlier, in data annotation task, workers need to choose one from several candidate labels which suits the task best. Intuitively, the more potential labels, the more difficult the problem will be. We assume the number of candidate labels is $k$. $e_j$ is adopted to represent the potential ability of worker $w_j$, and $d^i$ to represent the difficulty of task $t^i$. $\mathcal{E} = \{e_1, e_2, \ldots, e_m\}$ is chosen to denote the set of workers' abilities, and $\mathcal{D} = \{d^1, d^2, \ldots, d^n\}$ to denote the set of tasks' difficulties. We use a probability graph model to simulate the generation process of a single annotation $a_j^i$, as shown in *Fig. 1*. In the following sections we will analyse in detail the reason for the modeling of each step in *Fig. 1* and derive the calculation formulas which we rely on to establish our truth inference algorithm. According to *Fig. 1*, we can get the joint probability density formula as *Eq. 1*.

$$p(\mathcal{A}|\mathcal{D}, \mathcal{E}, \mathcal{L}) = p(\mathcal{P}|\mathcal{D}, \mathcal{E})p(\mathcal{A}|\mathcal{P}, \mathcal{L}) \tag{1}$$

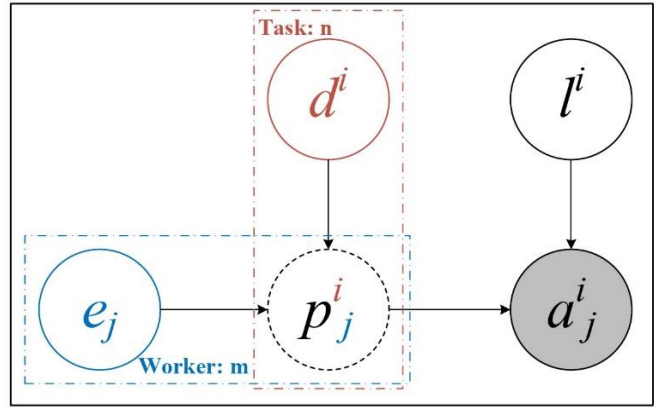

**Figure 1.** The probability graph model of proposed method. Above there are five elements: worker ability $e_j$, task difficult $d^i$, the probability $p_j^i$ indicts the confidence of worker's annotation $a_j^i$ being the same as ground truth $l^i$. The element in shaded circle ($a_j^i$) is the only variable we know, in dashed circle ($p_j^i$) is the intermediate variable, what we need to infer are the elements in the remaining three circles.

### 3.2. Worker-Task Interaction Model

Due to the open nature of the Internet, crowdsourcing workers may come from all corners around the world. They have different ethnicity, nationality, educational background, work skill, cognitive inclination, personal personality, *etc.* [19]. In order to highlight and study the different inherent performance of these workers during crowdsourcing, many worker modeling methods have been proposed, such as methods based on ability level (accuracy), methods based on ability domains, and methods based on confusion matrix, from simple to complex. [19] further classified workers to five classes based on their ability and behaviours.

Since the annotation matrix of a worker in real crowdsourcing scene is usually sparse, too complex models will suffer more from data lacking, so we employ the simplest and intuitive but practical ability model to represent the worker's ability. Each worker $w_j$ is associated to a decimal $e_j \in [0,1]$ indicating the ability to get task right. With reference to the work of [19] we pre-established four categories of workers based on their abilities and behaviours: *expert*, *normal* worker, *sloppy* worker and *spammer*. Expert and normal worker will do their best to get things right while sloppy workers tend to give out answers imprudently and there may be very few workers deliberately give out wrong answers for various reasons, trying to sabotage the current task.

Supposing you are taking an exam, besides your own knowledge and ability, the factors that determine your final score also include the quality of test paper, such as the difficulty of the test questions, the scope of investigation, the quality of the question description, and the length of the test duration, *etc.* Reasonable task design means a lot to the quality of crowdsourcing [6], but this is beyond the scope of this paper. We synthetically use a decimal $d^i \in [0,1]$ to quantify the difficulty of task $t^i$, from easy to difficult.

According to the experience in real life, a question may be too difficult for even experts to solve it well within limited time, identifying the authenticity of the porcelain in a picture, for example. On the other hand, even sloppy can easily answers the simplest problems, such as "*Which mountain is the highest in the world?*". Imagining what will happen if you encounter a question you don't know in an exam? picking an answer at random! The harder the question, the more difficult it is for workers to work out the correct answer, so the answers given tend to be chosen randomly. We use *Eq. 2* to express this trend:

$$p_j^i = f(d^i, e_j) = d^i \frac{1}{k} + (1-d^i)e_j^{d^i} \qquad (2)$$

here $p_j^i$ is the intermediate variable showing the possibility that the worker $w_j$ chooses what he thinks the correct answer is to task $t^i$ with $k$ choices. We choose *Eq. 2* because of the three important properties:

1. The function is concise, easy to calculate and understand. $f(d^i, e_j)$ can be seen as a competition between two factors, random selection and worker effort, difficulty ($d^i$, $(1-d^i)$) act as weights.
2. Fixing one of the two variables, the function $f(d^i, e_j)$ basically changes monotonously with the other variable, which is in line with our intuition. Some special cases are shown in *Fig. 2*.
3. In two extreme cases, the function value meets our original assumptions. $f(e_j|d^i=0)=1$ and $f(e_j|d^i=1)=\frac{1}{k}$ regardless of the value of $e_j$. Those mean that tasks that are too simple or too difficult are not enough to distinguish workers' ability difference.

When the difficulty of the task is 1 ($f(e_j|d^i=1)$), *Eq. 2* degenerates into the probability of random selection, in another extreme cases ($f(e_j|d^i=0)$), *Eq. 2* shows the all workers have full confidence to their answers. It can be seen from the *Fig. 2* that when the task difficulty is moderate (about 0.5), the distance between the confidence curves of the three types of workers is the largest, meaning that the task can distinguish workers of different abilities well at this time.

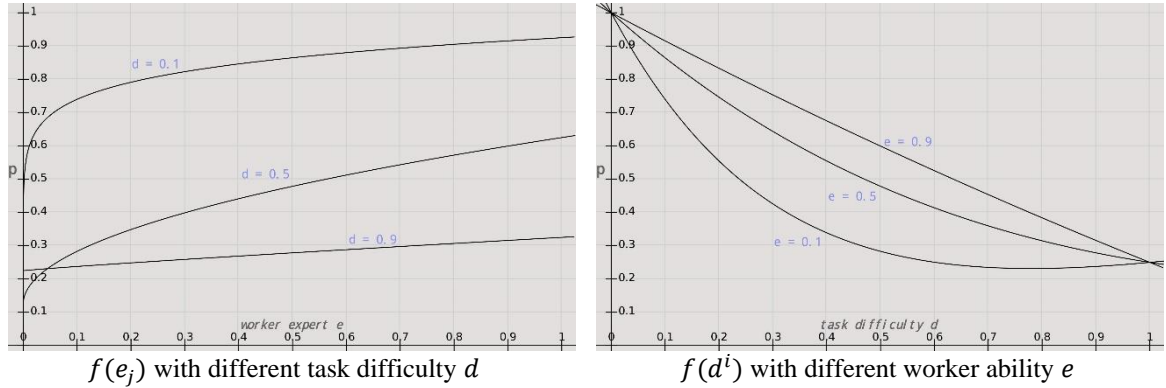

$f(e_j)$ with different task difficulty $d$      $f(d^i)$ with different worker ability $e$

**Figure 2.** The figure on the left shows the change curve of the answer quality on three typical difficulty tasks with the ability of the workers, and the right one shows the change curve of the answer quality given by the three typical workers with the difficulty of the task. Note that we set $k=4$ in above cases and the effective value of the horizontal coordinates is in the range of $[0,1]$.

Intuition teaches us that the higher the credibility of the annotation given by the worker $w_j$, the more likely his annotation is the same as the potential ground truth of task $t^i$. Combining $p_j^i$ and $l^i$, we can easily get the probability density function of the distribution of $a_j^i$ as *Eq. 3*.

$$\begin{aligned} f(a_j^i = l^i) &= p_j^i \\ f(a_j^i \neq l^i) &= \frac{(1-p_j^i)}{(k-1)} \end{aligned} \qquad (3)$$

We simply assume that all wrong options are equally confusing, so we use a simple equalization process to represent the probability distribution density of the wrong answers. Although the use of more accurate models such as confusion matrix will be closer to the real situation, the improvement they bring is limited when the amount of data and information is relatively scarce [20].

$$f(a_j^i = l^i) = p_j^i = \frac{d^i}{k} + (1 - d^i)e_j^{d^i} \approx \frac{1}{k} \quad s.t. \ d^i \approx 1 \tag{4}$$

### 3.3. Truth Inference Algorithm

In *Fig. 1*, there are four elements, of which only annotation $a_j^i$ is what can be observed, how to infer the other three elements from the mere information becomes a problem. Generally speaking, when the value of $a_j^i$ is given, $l^i$ and $p_j^i$ cannot be independent of each other. However, $p_j^i$ is a hidden variable, and we have no idea about its value. In this case, $e_j$ and $d^i$ are independent of each other (see *Eq. 5, 6, 7*). These analyses are in line with our life experience. Therefore, when the value of $a_j^i$ is observed, the three element we are concerned with: task label, worker ability, and task difficulty have some correlations. $l^i$ depends on $d^i$ and $e_j$ both, while the later two are mutually independent.

Joint probability formula:

$$P(p_j^i, l^i, a_j^i) = P(p_j^i)P(l^i|p_j^i)P(a_j^i|p_j^i, l^i) \tag{5}$$

Bayesian hypothesis:

$$P(p_j^i, l^i, a_j^i) = P(p_j^i)P(l^i)P(a_j^i|p_j^i, l^i) \tag{6}$$

Compare above we have:

$$P(l^i|p_j^i) = P(l^i) \rightarrow P(l^i, p_j^i) = P(l^i)P(p_j^i) \tag{7}$$

Luckily, a task will usually be annotated by multiple workers and a worker will annotate a set of tasks. We assume that the ability of a worker is stable for a certain period of time, and the difficulty of the task is the attribute of the task itself, so we can transfer the ability of a worker on $t^i$ task to $t^j$ task, and the difficulty of a task inferred from the performance of worker $w_p$ can also be used to predict the performance of worker $w_q$.

We propose an iterative calculation method based on the *EM* algorithm to update the values of $\mathcal{E} = \{e_1, e_2, ..., e_m\}$, $\mathcal{D} = \{d^1, d^2, ..., d^n\}$ and $\mathcal{L} = \{l^1, l^2, ..., l^n\}$. In each iteration, we use the current values of two of the three sets to update the other one, one iteration ends when the values of all three sets are updated once. In the related work section, we mentioned that the quality of the initial value is very important for the iterative algorithm. Therefore, in the following steps, we will explain in detail how to set the initial value at the start-up and how to update the above three sets.

Generally speaking, researchers will use a certain strategy to initialize the worker and task model, like random initialization or uniform initialization, and then derive the truth label set. Few people will assume the ground truths of the task to start the iteration, because this is contrary to our common sense. So, in each iteration we always use $\mathcal{E}$ and $\mathcal{D}$ to update $\mathcal{L}$ first, then $\mathcal{E}$ or $\mathcal{D}$.

#### 3.3.1. Parameter Initialization

The work of [20] has summarized that when the amount of data is sufficient, simple models such as *MV* or *WMV* can also achieve good results, and have more stable performance and less calculation than complex models. In other words, simple algorithms such as *MV* or *WMV* tend not to be too bad and fluctuate less. So, we use simple *MV* as example to explain how to set the initial value of our model, it contains the two steps:

*Count*: First, for each task, we count all the annotations it receives, accumulate the number of counts for each option, and the option with the highest score is treats as the label of the task. Then we use the error rate as the initial difficulty of the task.

*Weight*: After obtaining the labels of all tasks, we in turn count all the annotations given by each worker, and use the correct rate as the initial value of the worker's ability.

$$d^i = 1 - \frac{\sum_{a_j^i \in A^i} \mathcal{I}(a_j^i = \hat{l^i})}{|A^i|} \tag{8}$$

$$e_j = \frac{\sum_{a_j^i \in A_j} \mathcal{I}(a_j^i = \hat{l^i})}{|A_j|} \tag{9}$$

In *Eq. 8&9*, $\mathcal{I}()$ is the indicator function, $\mathcal{I}(True) = 1$, $\mathcal{I}(False) = 0$, the hat of $\hat{l^i}$ denotes it is an approximate of the latent ground truth. If you want to get more accurate initialization parameters, or the number of annotations $r \leq k$, you can also use other iteration-based methods, such as *WMV* to initialize the parameters to avoid the embarrassing situations where there is no option to win. The main difference in *WMV* is that in the *Count* stage, when we count the score of each option, we will multiply the weight of each worker's annotation, that is, the accuracy of the worker in the current iteration. Then iteratively perform the *Count* and *Weight* steps until it stabilizes. The error rate is treated as task difficulty and correct rate as the worker ability.

In the following update process, different from the previous meaning, we use superscript on a set to mark the iteration rounds, $\mathcal{L}^{i+1} = f(\mathcal{D}^i, \mathcal{E}^i)$ means we use the task difficulty set $\mathcal{D}$ and worker ability set $\mathcal{E}$ obtained in round $i$ to update the task label set in round $i + 1$, for example.

### 3.3.2. Parameter Initialization

The term of the unobserved variable is called "latent variable". Let $\mathcal{X}$ denote the set of observed variables, $\mathcal{Z}$ denote the set of latent variables, $\Theta$ denote the model parameters, and $LL()$ denote the log-likelihood function. If we want to do the maximum likelihood estimation of $\Theta$, we have *Eq. 10*. Since $\mathcal{Z}$ is unknown, we can only maximize the logarithmic marginal likelihood of the observed variables $\mathcal{X}$ by calculating the expectation of $\mathcal{Z}$, as shown in *Eq. 11*.

$$LL(\Theta|\mathcal{X}, \mathcal{Z}) = \ln P(\mathcal{X}, \mathcal{Z}|\Theta) \tag{10}$$

$$LL(\Theta|\mathcal{X}) = \ln P(\mathcal{X}|\Theta) = \ln \sum_{\mathcal{Z}} P(\mathcal{X}, \mathcal{Z}|\Theta) \tag{11}$$

In E-step, we need to infer the distribution of latent variables $\mathcal{Z}$ under current model parameters $P(\mathcal{Z}|\mathcal{X}, \Theta^i)$, and calculate the expectation of the log-likelihood function $LL(\Theta|\mathcal{X}, \mathcal{Z})$ for $\mathcal{Z}$, as in *Eq. 12*.

$$Q(\Theta|\Theta^i) = \mathbb{E}_{\mathcal{Z}|\mathcal{X}, \Theta^i} LL(\Theta|\mathcal{X}, \mathcal{Z}) \tag{12}$$

In M-step, we can solve *Eq. 13* to update the model parameters $\Theta^{i+1}$ to maximize the expectation of the likelihood function $Q(\Theta|\Theta^i)$.

$$\Theta^{i+1} = \max_{\Theta} Q(\Theta|\Theta^i) \tag{13}$$

**Updating $\mathcal{L}$:**
We first use $(\mathcal{D}^i, \mathcal{E}^i) \to \mathcal{P}^i$ to get $\mathcal{P}^i$, then take $(\mathcal{P}^i, \mathcal{A}) \to \mathcal{L}^{i+1}$ to get the distribution of $\mathcal{L}$ in $(i+1)$ turn. Those can be seen as the E-step.

**Updating $\mathcal{D}$ & $\mathcal{E}$:**
When we have the new distribution of $\mathcal{L}^{i+1}$, we can calculate the difficulty of all tasks ($\mathcal{D}^{i+1}$) by *Eq. 8*, showing as $(\mathcal{L}^{i+1}, \mathcal{A}) \to \mathcal{D}^{i+1}$. But how to update the worker ability is a bit different from *Eq. 9*. Remember the core idea of this paper, more than ten years of examination experience tells us that questions with different difficulty have different ability to reflect the ability of workers, we also designed the worker-task interaction function as *Eq. 2*, based on this concept. So in $(\mathcal{L}^{i+1}, \mathcal{D}^{i+1}) \to \mathcal{E}^{i+1}$, we modify *Eq. 9* to *Eq. 18*. It is not difficult to find that we also weighted the contribution of different tasks to the ability of a worker, just like weighting the contribution of different workers' annotations to a task label. *Eq. 14* shows the probability difference between two workers subject to task difficulty and *Eq. 15* is the derivative. From *Eq. 16 & 17* we can see that tasks with moderate difficulty ($d \approx 0.4$) have the highest distinguishing ability, so we should reduce the interference of extreme tasks on deriving the ability of workers. Finally, the weight is set as *Eq. 19*.

$$F(d) = f(e_1, d) - f(e_2, d) = (1-d)(e_1{}^d - e_2{}^d) \; s.t. \; 1 > e_1 > e_2 > 0 \tag{14}$$

$$F'(d) = e_2{}^d - e_1{}^d + (1-x)(e_1{}^d \ln e_1 - e_2{}^d \ln e_2) \tag{15}$$

$$\mathbb{E}(e_1 - e_2) = \mathbb{E}((e_2 + \frac{1-e_2}{2}) - e_2) = \mathbb{E}(\frac{1-e_2}{2}) = 0.25 \; s.t. \; 1 > e_1 > e_2 > 0 \tag{16}$$

$$F'(0.4) \approx 0 \; s.t. \; \mathbb{E}(e_1 - e_2) = 0.25 \tag{17}$$

$$e_j = \frac{\sum_{a_j^i \in A_j} \phi(d^i) \mathcal{I}(a_j^i = \hat{l}^i)}{\sum_{a_j^i \in A_j} \phi(d^i)} \tag{18}$$

$$\phi(d^i) = \begin{cases} 1 - \frac{d^i - 0.4}{0.6}, & d^i \geq 0.4 \\ 1 - \frac{0.4 - d^i}{0.4}, & d^i < 0.4 \end{cases} \tag{19}$$

We iteratively perform the above steps until the predetermined number of iteration is reached or the parameters no longer change. The final outputted $\mathcal{L}$ is the result of the tasks we inferred.

### 3.4. Task Assignment

Task assignment strategy is affected by many factors [5], such as platform task allocation method, modelling method, whether the budget is sufficient, the average number of workers' annotations and the average number of repeated annotations on a task, *etc.* Since our model needs to understand the task and worker information at the same time, this requires more information to build the model. At the same time, we have carried out a two-way correction on the mutual derivation between task difficulty and worker ability, which can reduce the impact of random assignment. Therefore, when the budget is not sufficient ($r < 3$), we recommend using random assignment. When the budget is relatively sufficient ($r > 5$), we can first use random assignment to build the task-worker model, then sort the workers and tasks separately, and then use the remaining budget for matching assignment.

```
Algorithm 1: Worker-Task Interaction Model
  Input: task set 𝒯, worker set 𝒲, budget
  Output: annotation set 𝒜, task difficulty set 𝒟,
          worker ability set ℰ, inferred label set 𝓛̂
1 while (r < 3) do
2  │  Random assign 𝒯 to 𝒲 to fill 𝒜;
3 end
4 Initialize 𝒟⁰ and ℰ₀ using Eq. 8, 18, 19;
5 while not converged do
6  │  Update 𝓛^{i+1}: (𝒟^i, ℰ_i) → 𝓛^{i+1} using Eq. 12;
7  │  Update 𝒟^{i+1} and ℰ^{i+1} using Eq. 13;
8 end
9 Sort 𝒯 and 𝒲 for remaining assignment;
10 while (budget > 0) do
11 │  Assign tasks in corresponding order to update 𝒜;
12 end
13 Re-derive 𝒟, ℰ, 𝓛 and output the final 𝓛;
```

## 4. EXPERIMENTS AND ANALYSIS

Since we have proposed a fresh worker-task interaction model (note as *WTIM*), we intend to prove the effectiveness and practicability of our algorithm in three steps. We first verify the effectiveness of our model under ideal conditions through a series of simulation experiments, then we run our algorithm on two real crowdsourcing data sets to test its practical performance. Finally, we hired colleagues around us to complete a special test to test the correctness of our hypothesis. In the real world, the actual model of crowdsourcing workers is affected by time, place, task, platform, workers' own situation and other random factors, and no worker model is confirmed to be universally correct [21]. Therefore, we believe that the method proposed by us is irreplaceable in some scenarios.

For single-choice tasks, the only evaluation indicator used is the correct rate or accuracy of the final answer. Since our truth inference method is based on an iterative calculation, in addition to the most common baseline algorithm *MV* and *WMV* in crowdsourcing, we also compared some classic iterative-based truth inference algorithms. Their brief introduction is as follows:
1. *MV*: *MV* directly uses majority vote to integrate annotations, without modeling tasks and workers.
2. *WMV*: *WMV* models each worker a weight when aggregating answers, that is, their accuracy. By iteratively updating task labels and worker ability, *WMV* can finally output a much more accurate result than *MV*, many other methods such as *ZC* [22] are based on *WMV*.
3. *ZC*: *ZC* [22] adopts a probabilistic graphical model to model the decision-making process of workers. In addition, its worker model uses the simplest accuracy and does not model the task. It also uses the *EM* algorithm to iteratively update the model parameters and task results.
4. *DS*: *DS* [13] is a classic worker model. It uses a confusion matrix to simulate the different tendencies of workers when choosing answers. The sum of each row or column in the matrix is 1, and the value of the $i,j$-th element represents the probability that the worker chooses item $j$ when the true value is item $i$.
5. *GLAD*: *GLAD* [15] extend *WMV* in task model. Instead of treating tasks equally, *GLAD* gives each task a difficulty $d^i \in (0, +\infty)$ (the bigger, the easier), it further models workers by accuracy and the annotation distribution as $Pr(a^i = l^i | d^i, e_j) = 1/(1 + \exp^{(-d^i * e_j)})$, some way the same like us.

### 4.1. Simulation Experiment Setup

In the simulation experiment, we simulate the random process of a certain group of workers annotating a certain set of tasks according to certain rules. We first determine some parameters about the task, such as the number of options $k$, then we generate tasks following the setting in *Tab. 1*. For workers, we refer to the work of [19] and set up four groups of workers with different behaviour models. Their specific settings and ratios are shown in *Tab. 2*. We generated a total of $n$ tasks and $m$ workers, and their difficulties or abilities randomly fluctuate within the ability fluctuation range of their respective groups in two tables.

**Table 1**. Task classification and corresponding difficulty range and proportion of the simulated tasks. $k$ is the number of options in the crowdsourcing task.

| Worker Type | Difficulty Baseline | Fluctuation | Proportion |
|---|---|---|---|
| Hard | 0.85 | ±5% | 10% |
| Medium | (1 + 1/k)/2 | ±10% | 60% |
| Easy | 0.15 | ±5% | 30% |

**Table 2**. Worker classification and corresponding ability range and proportion of the simulated workers. $k$ is the number of options in the crowdsourcing task.

| Worker Type | Ability Baseline | Fluctuation | Proportion |
|---|---|---|---|
| Expert | 0.85 | ±5% | 30% |
| Normal | (1 + 1/k)/2 | ±10% | 40% |
| Sloppy | 1/k | +10% | 20% |
| Spammer | 0.15 | ±5% | 10% |

In crowdsourcing research, a lot of works focus on binary classification task, or True-False task, such assumptions are relatively simple but without loss of generality, they will also claim that multi-classification task can also be obtained by combining several binary classification tasks. An interesting property of the binary classification task is that the accuracy of random guessing will not deviate far from 0.5, so we can easily classify workers and correct the answers from unqualified workers. However, in practice, many tasks have multiple alternative options, and the native binary classification algorithm is not easy to handle them well, therefore, we set up two kinds of simulation experiments, $k = 2$ and $k = 4$ respectively. According to *Tab. 1* and *Tab. 2*, we have simulated $n = 100$ tasks and $m = 10$ workers, and each task is annotated $r = 5$ times on average[1], which means that each worker will annotate 50 tasks on average.

### 4.2. Simulation Results & Analysis

---

[1] As suggested in [15, 17, 20], the recommended value of $r$ ranges in [3 ~ 7], and for single choice task with four options, the results keep stable for $r \geq 5$.

The results of the simulation experiment are recorded in *Tab. 3*. We roughly recorded the running time of each algorithm and the accuracy of the results, and the best results are marked in bold. It can be seen from the table that the results of the methods (*WMV, ZC, DS, GLAD, WTIM*) with modeling of workers and/or tasks are far better than *MV*, and the differences between them are not obvious. Note that *WTIM* does not know the specific worker and task model in the truth inference stage. Because the model of our algorithm is the closest to the simulated worker-task model, we have achieved the best accuracy, which shows that our algorithm can indeed work well in certain scenarios. In addition, note that algorithms based on iteration and probabilistic graph models (*GLAD, WTIM*) often take a long time to stabilize, therefore, in actual crowdsourcing, quality, budget, and delay often need to be considered comprehensively [4].

**Table 3**. The performance of our method and various comparison methods on the simulated datasets. We calculate the algorithm accuracy and iteration time respectively. The result of the winning algorithm is highlighted in bold.

| Dataset & KPI | MV | | WMV | | ZC | | DS | | GLAD | | WTIM | |
|---|---|---|---|---|---|---|---|---|---|---|---|---|
| | ACC | Time | ACC | Time | ACC | Time | ACC | Time | ACC | Time | ACC | Time |
| Simulation data (k = 4) | 52% | **<1s** | 63% | <1s | 62% | <1s | 64% | <1s | 66s | 325s | **69%** | 558s |

**Table 4**. The performance of our method and various comparison methods on the two selected real world data sets. We calculate the algorithm accuracy and iteration time respectively. The result of the winning algorithm is highlighted in bold.

| Dataset & KPI | MV | | WMV | | ZC | | DS | | GLAD | | WTIM | |
|---|---|---|---|---|---|---|---|---|---|---|---|---|
| | ACC | Time | ACC | Time | ACC | Time | ACC | Time | ACC | Time | ACC | Time |
| BM(k = 2) | 69.6% | **<1s** | 69.6% | <1s | 68.9% | <1s | 69.5% | 1s | **70.1%** | 1154s | 69.9% | 2765s |
| AC(k = 4) | 75.97% | **<1s** | 76.27% | <1s | 75.97% | 2s | 76.57% | 3s | 76.87% | 3866s | **77.17%** | 4529s |

### 4.3. Dataset Experiment Setup

Thanks to the work of Sheng [17], Barzan Mozafari and Zheng [20], we have found some useful real world crowdsourcing dataset. Those datasets are available at here[2]. They are all about single-choice tasks, in which workers are asked to choose one label from four candidates. Brief introductions of the two data sets are as follows:
1. Barzan Mozafari's (note as *BM*) dataset: It is a binary classification task data set that contains a total of 1000 tasks to which AMT workers provide a total of 5000 annotations. All of them have ground truth labels and each task has 5 annotations on average.
2. Adult Content (note as *AC*) dataset: It contains about $100k$ annotations assigned by AMT workers to identify the adult level of the website, choosing from four classes, *G* (General Audience), *P* (Parental Guidance), *R* (Restricted) and *X* (Porn). Only 333 tasks annotated by workers have ground truth labels, so we only count the results of this part. Each task has around 10 annotations on average.

### 4.4. Dataset Results & Analysis

We run our algorithm and comparison methods on these two data sets until convergence, and their results are statistically in *Tab. 4*. As can be seen from the table, the difference between the various methods is not obvious for these two data sets, which may be explained by that the workers to complete the two set of tasks are all qualified and stable. For data set *AC*, our results and [20]'s are basically consistent, and the numerical differences may come from the inconsistency of the statistical standards on the number of annotated samples, we compare files containing gold truth and files containing labels, eventually only get 333 annotated tasks with ground truth label, the numbers of correct tasks in these true inference algorithms are all about 255. *WTIM* and *GLAD* are neck and neck in accuracy, but they both are the most time-consuming algorithms.

---
[2] https://github.com/ipeirotis/Get-Another-Label

Another fact is that the results of *WMV* and *MV* are basically consistent on these two datasets. In other words, each task is well completed in the limit of workers' capabilities, therefore the improvement carried by various true inference algorithm is not obvious. On the other hand, this also shows that our method is feasible on real dataset, at least it does not deviate from the result of the mainstream algorithm too far.

### 4.5. Real World Test

We specially designed a quiz consisting of 20 four-option single-choice questions, including elementary mathematics content, computer science content and geography content, and their proportions are shown in *Tab. 1*. For students in the School of Computer Science, we think that the difficulty of these three parts is gradually increasing. We invite ten classmates to complete this quiz for free and give out scores and rankings. Interestingly, almost all the classmates successfully answered the first two parts of the quiz, and their scores on geography questions far exceeded the random choice $(12/(2*10) > 1/4)$. From this point of view, in real crowdsourcing applications, in addition to task assignment and truth inference algorithm, how to attract higher-quality workers and maintain their enthusiasm by good task design and incentive mechanisms are also important factors to improve crowdsourcing quality.

## 5. CONCLUSIONS

In this paper, we extracted the model of worker-task interaction in the process of crowdsourcing from actual exams, and transformed it into a probabilistic graphical model. We believe that the difficulty of the task and the ability of the workers will both affect the quality of the final task, and there is an interactive relationship between the two. Through simulation experiment and experiments on real datasets, we have proved that our algorithm is better than existing algorithms to a certain extent, and has irreplaceable advantages in some characteristic scenarios. Finally, experiments and quiz show that, in addition to the later data mining efforts, the early task design, incentive settings design and other humanistic efforts are equally important in crowdsourcing, in fact, good raw data can make truth inference algorithms less important.